\RequirePackage{rotating}
\documentclass[preprint]{imsart}

\usepackage{blkarray}
\RequirePackage{amsthm,amsmath,amsfonts,amssymb,mathtools,bm,mathabx,tikz}
\RequirePackage[authoryear]{natbib}%% uncomment this for author-year citations
\RequirePackage[colorlinks,citecolor=blue,urlcolor=blue]{hyperref}%% uncomment this for coloring bibliography citations and linked URLs
\RequirePackage{graphicx}%% uncomment this for including figures
\usepackage[boxed, ruled, lined, linesnumbered]{algorithm2e}
\usepackage[nameinlink]{cleveref}
\usepackage{multirow}
\usepackage{color}
\usepackage{listings}
\usepackage{caption}
\usepackage{tabularx}
\usepackage{float}
\usepackage{bbm}
\usepackage{enumerate}
\startlocaldefs
\theoremstyle{plain}

\newtheorem{prop}{Proposition}
\newtheorem{cor}{Corollary}
\theoremstyle{remark}
\newtheorem{ass}{Assumption}

\newtheorem{defn}{Definition}
\newtheorem{property}{Property}

\newtheorem{rem}{Remark}
\newenvironment{ass*}
 {\expandafter\def\expandafter\theass\expandafter{\theass*}\ass}
 {\endass}
\renewcommand{\sectionautorefname}{Section}

\Crefname{ass}{Assumption}{Assumptions}
\Crefname{prop}{Proposition}{Propositions}
\Crefname{section}{Section}{Sections}
\Crefname{appendix}{Appendix}{Appendices}
\Crefname{cor}{Corollary}{Corollaries}
\Crefname{defn}{Definition}{Definitions}
\Crefname{enumi}{Condition}{Conditions}
\Crefname{property}{Property}{Properties}
\Crefname{rem}{Remark}{Remarks}

\DeclareMathOperator*{\argmin}{arg min}

\newcommand{\trace}{\operatorname{tr}}
 
\endlocaldefs

\begin{document}

\begin{frontmatter}
%%%%%%%%%%%%%%%%%%%%%%%%%%%%%%%%%%%%%%%%%%%%%%
%%                                          %%
%% Enter the title of your article here     %%
%%                                          %%
%%%%%%%%%%%%%%%%%%%%%%%%%%%%%%%%%%%%%%%%%%%%%%
%\title{Non-Parametric Ridge Recovery of Atomic Columns in Transmission Electron Microscopy Videos} %Image Sequences

 \title{Dynamic Atomic Column Detection in Transmission Electron Microscopy Videos via Ridge Estimation}
%\title{A sample article title with some additional note\thanksref{T1}}
%\runtitle{TEM Image Series Ridge Recovery}
\runtitle{Ridge Recovery for TEM Sequences}
%\thankstext{T1}{A sample of additional note to the title.}

\begin{aug}
%%%%%%%%%%%%%%%%%%%%%%%%%%%%%%%%%%%%%%%%%%%%%%%
%% Only one address is permitted per author. %%
%% Only division, organization and e-mail is %%
%% included in the address.                  %%
%% Additional information can be included in %%
%% the Acknowledgments section if necessary. %%
%% ORCID can be inserted by command:         %%
%% \orcid{0000-0000-0000-0000}               %%
%%%%%%%%%%%%%%%%%%%%%%%%%%%%%%%%%%%%%%%%%%%%%%%
\author[A]{\fnms{Yuchen}~\snm{Xu}\ead[label=e1]{yx439@cornell.edu}},
\author[A]{\fnms{Andrew M.}~\snm{Thomas}\ead[label=e2]{amt269@cornell.edu}},
\author[B]{\fnms{Peter A.}~\snm{Crozier}}\ead[label=e3]{crozier@asu.edu},
\and
\author[A]{\fnms{David S.}~\snm{Matteson}\ead[label=e4]{matteson@cornell.edu}}
%%%%%%%%%%%%%%%%%%%%%%%%%%%%%%%%%%%%%%%%%%%%%%
%% Addresses                                %%
%%%%%%%%%%%%%%%%%%%%%%%%%%%%%%%%%%%%%%%%%%%%%%
\address[A]{Department of Statistics and Data Science, Cornell University\printead[presep={,\ }]{e1,e2,e4}}
\address[B]{School for Engineering of Matter, Transport and Energy, Arizona State University\printead[presep={,\ }]{e3}}
\end{aug}

\begin{abstract}
Ridge detection is a classical tool to extract curvilinear features in image processing. As such, it has great promise in applications to material science problems; specifically, for trend filtering relatively stable atom-shaped objects in image sequences, such as Transmission Electron Microscopy (TEM) videos. Standard analysis of TEM videos is limited to frame-by-frame object recognition. We instead harness temporal correlation across frames through simultaneous analysis of long image sequences, specified as a spatio-temporal image tensor. We define new ridge detection algorithms to non-parametrically estimate explicit trajectories of atomic-level object locations as a continuous function of time. Our approach is specially tailored to handle temporal analysis of objects that seemingly stochastically disappear and subsequently reappear throughout a sequence. We  demonstrate that the proposed method is highly effective and efficient in simulation scenarios, and delivers notable performance improvements in TEM experiments compared to other material science benchmarks.
\end{abstract}

% \begin{keyword}[class=MSC]
% \kwd{62H15}
% \kwd{62G05}
% %\kwd[; secondary ]{???}
% \end{keyword}

%\kwd{continuous curve}

% \kwd{vector autoregression}
% \kwd{Markov chain}
% \kwd{Wald test}
% \kwd{dimension reduction}

\end{frontmatter}
%%%%%%%%%%%%%%%%%%%%%%%%%%%%%%%%%%%%%%%%%%%%%%
%% Please use \tableofcontents for articles %%
%% with 50 pages and more                   %%
%%%%%%%%%%%%%%%%%%%%%%%%%%%%%%%%%%%%%%%%%%%%%%
%\tableofcontents

%%%%%%%%%%%%%%%%%%%%%%%%%%%%%%%%%%%%%%%%%%%%%%
%%%% Main text entry area:

\section{Introduction}\label{sec:intro}
\begin{figure}[htb]
    \captionbox{The sketch of a time-resolved \emph{in situ} TEM video of a $\mbox{CeO}_2$ nanoparticle with temporal resolution 2.5 milliseconds. Here, the $x$ and $y$ axes identify the spatial coordinates within single TEM image frames, and the $t$ axis represents the dimension of time. Every single frame visualizes the physical nanoparticles' model as the gray-scaled projection image, and the whiter regions are the places where atoms are stacked to form atomic columns perpendicular to the page. A zoomed view for one of the atomic columns is provided as \autoref{fig:blob_ridge}.\label{fig:tem_eg}}[.5\textwidth]
        {\includegraphics[width=.5\textwidth]{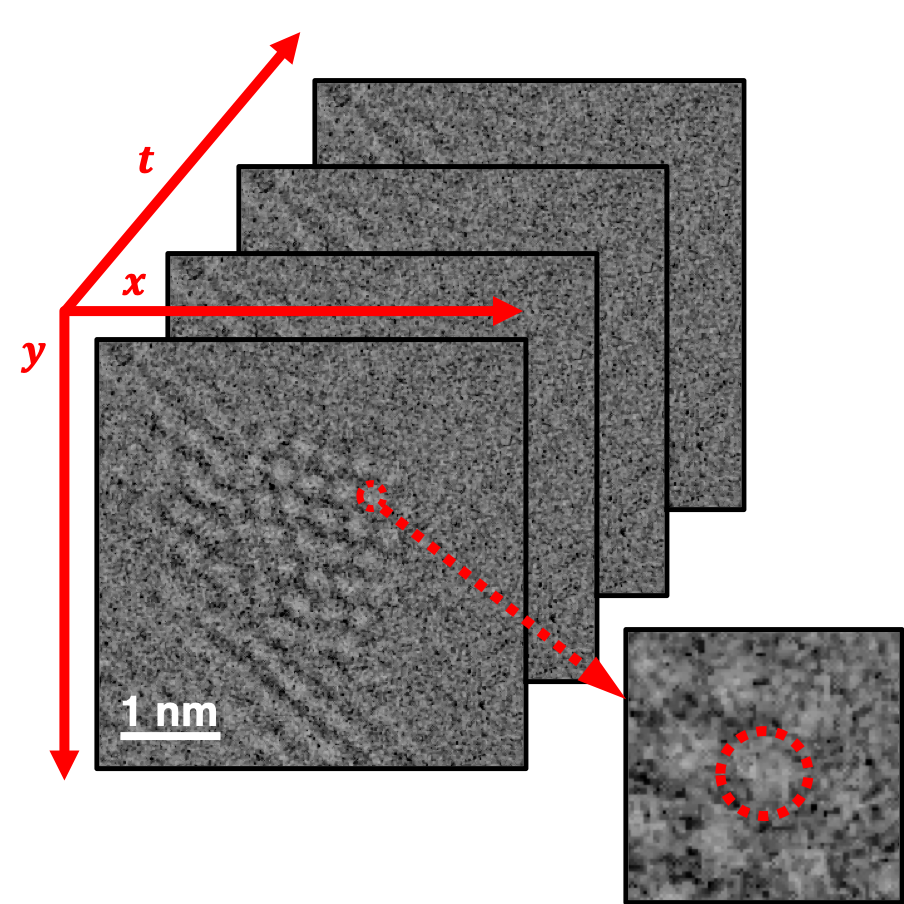}}\hspace{.05\textwidth}%
    \captionbox{The illustration of a generalized ridge curve $\gamma \subset \mathbb{R}^3$ as a trajectory function of time $t$ for a selected atomic column after negating the TEM video. The negation is supported as in \autoref{rem:def}. In the example sequence, the atomic column experiences various dynamics such as shrinking and expanding radius, decreasing and increasing contrasts, and total degeneration to a single point in the middle of the series. In addition, the outer contour of the atomic column forms a tube-shaped object that stretches temporally.
    % It also consists of shrinking radius, varying contrasts and occasional degeneration (absence).
    \label{fig:blob_ridge}}[.45\textwidth]
        {\includegraphics[width=.4\textwidth]{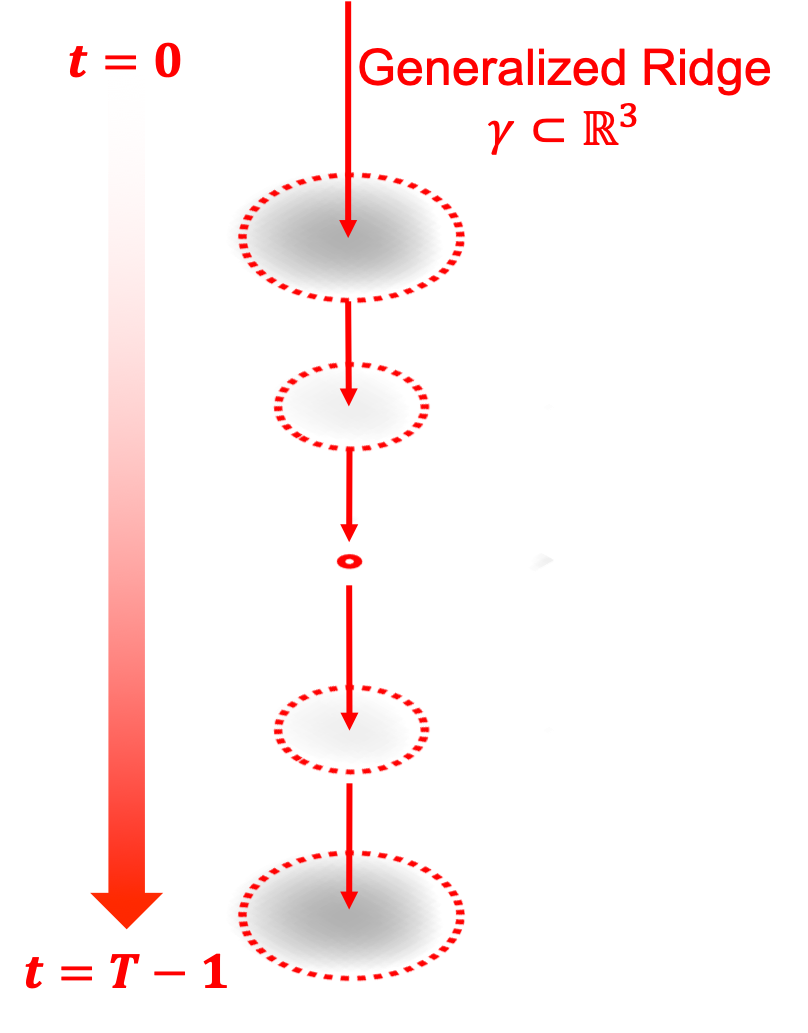}}
\end{figure}

Transmission electron microscopy (TEM) is an essential tool for studying materials at the atomic level. Major technical advancements in TEM have improved not only temporal resolution but also sensitivity for atomic structure detection \citep{ruskin2013a,faruqi2018,levin2021}. TEM images are known to suffer critical degradation in signal-to-noise ratios (SNR) \citep{lawrence2020,lawrence2021,vincent2021} 
due to the scientific demands for instantaneous analysis coupled with the limited equipment capacity of electron flux \citep{egerton2004,egerton2013,egerton2019}.
\autoref{fig:tem_eg} is an illustration of a real TEM video with high temporal resolution, where the severe SNR challenge can be read from the example frame.
Hence, improved analysis algorithms for TEM images and videos are in continuous demand, and incubated at varied focus and strength \citep{lawrence2020}. Yet, important weaknesses still remain and  accompanying data-intensive methodology is urgently needed.

In analysis of gray-scale TEM images, the most essential task is to estimate (detect) nanoparticle structure and summarize (extract) important atomic features to quantify and understand physical atomic dynamics, especially for \emph{in situ} experiments with catalysts. 
Fitting Gaussian models is the classical benchmark method and enjoys its popularity due to its straightforward interpretations. This requires various estimation routines including nonlinear least-square Gaussian Peak Fitting (GPF, \citealt{levin2020}) and Atomap (AM, \citealt{nord2017}). 
However, it is restrictive in terms of elliptical shape assumptions, and it is extremely computationally sensitive to both initialization as well as the SNR conditions.
%
%Being the classical benchmark method and enjoying its popularity due to straightforward interpretations, fitting Gaussian models, with various estimation routines including nonlinear least-square Gaussian Peak Fitting (GPF, \citealt{levin2020}) and Atomap (AM, \citealt{nord2017}), is restrictive with respect to elliptical shape assumptions and extremely sensitive to initialization as well as the SNR conditions at the computation stage. 
%
\cite{manzorro2022} recently introduced a blob detection (BD) approach tailored for TEM images. Their algorithm drops the explicit shape constraints and copes with the severe SNR challenge. Meanwhile, limitations remain, especially when methods are also needed to distinguish absence of atomic columns from vacuum background. \citep{thomas2022} proposed a topological data analysis (TDA) method to verify such differences using a hypothesis testing schema. The algorithm also requires less restrictive graphical assumptions. Some machine learning algorithms (e.g., \citealt{lin2021}) have also been developed, but the models therein are usually trained given very narrow SNR levels and current use cases lack generalizability.

Since most of these analyses have limited focus on individual frames, the assumptions ranging from graphical shapes to nanoparticles' intact lattice structures are usually necessary. Meanwhile these assumptions may frequently be violated in real experiments due to the atoms' volatile movements. On the other hand, the raw TEM data do not seem to be fully exploited.
% For instance, if we include the temporal dimension of TEM videos for collective analysis, most of these imposed assumptions may already be implicitly self-contained.
% Many algorithms extract the locations of atomic columns with prior boundary information (e.g., \citealt{levin2020,manzorro2022}), indicating the atomic columns' spatial stability to certain extent.
In this paper, we introduce a novel point of view on the spatio-temporal TEM image tensor (\autoref{fig:tem_eg}), and propose a generalized ridge detection (RD) algorithm that makes use of the rarely accounted for temporal correlation. Our ridge formulation is naturally motivated and justified by stacking the same atomic column's BD output blobs temporally to form a tube; see \autoref{fig:blob_ridge}. Note that such tubes in raw TEM videos usually correspond to valleys instead of ridges as the pixel intensities of atomic columns have smaller values than those of neighboring regions according to TEM mechanisms, hence negation is needed prior to our ridge processing.

Some of the existing milestone algorithms may utilize the TEM history only for denoising purposes during preprocessing. Our proposed ridge detection approach treats such information differently. It not only employs the data structure more profoundly to interpolate the curved cylinder of temporally-piled atomic columns, but also maintains sufficient independence between frames instead of completely smoothing out local features.

% If, however, certain dramatic dynamics occur in nano-particles' structures, the method may need additional uncertainty measures to compensate for the potential counterproductive effect.

% The assessment and justification therein against other competing algorithms address the strength of incorporating temporally correlated information when dealing with spatial objects.

% ridge detection and applications
Regarding image processing, \cite{lindeberg1996} first introduced the scale-space detection operator to adaptively account for lines' width. Since then, some state-of-art methodologies of ridge detection have been developed, such as \texttt{meijering} \citep{meijering2004}, \texttt{sato} \citep{sato1998}, \texttt{frangi} \citep{frangi1998}, and \texttt{hessian} \citep{ng2015}. Indeed, these algorithms reassign each pixel with a score of it being on a ridge, and reproduce the image with denoised and intensified curvilinear patterns.  The image processing python package \texttt{scikit-image} \citep{vanderwalt2014} offers a collection of the above four implementations. Subsequent works proposed different types of filtering measures and/or threshold schema with various tools. For example, \cite{norgard2013} introduced the combinatorial Jacobi sets to guarantee valid global structures, \cite{lopez-molina2015} used the Anisotropic Gaussian Kernel (AGK) to improve sensitivity and robustness, and \cite{reisenhofer2019} combined the contrast-invariant phase congruency measure with $\alpha$-molecules to refine local features given limited samples. For comprehensive reviews on ridge detection and enhancement algorithms, see \cite{shokouh2021,alhasson2021}, and references therein. Ridge detection has many applications including medical images analysis \citep{lopez1999}, fingerprint enhancement \citep{lindeberg2000}, autonomous navigation \citep{beyeler2014}, signal processing \citep{colominas2020,laurent2022}, and geology \citep{pham2021}, to name a few.

% different field of analysis
The related Lagrangian coherent structures (LCSs, \citealt{haller2000,haller2002}) methods are regarded as a ridge variant in mechanical engineering and dynamic systems \citep{mathur2007,senatore2011,schindler2012}; they aim to separate different fluid behaviors in finite-time Lyapunov exponent fields \citep{kasten2009}. Computationally, they are usually filtered using numerical differential equation solvers; see \cite{ameli2014} for examples.

% uncertainty quantification for multivariate data

% summary
%The literature review shows that the image processing community primarily focuses on 
Overall, the primary focus of the image processing community has been  
establishing filtered pixel-wise ridge-likelihood images for enhanced visual inspection, while the mechanical engineering community focuses its efforts on approximating ridges using discretized numerical solvers. Hence, analytical tools with the combination of ridge detection and continuous characterization are needed. The breakthrough of this work is to introduce innovative methodology to non-parametrically extract the continuous trajectories of atomic columns in TEM videos, with the tolerance of occasional degeneration throughout the sequence.
In practice, we refer to the idea in some existing methods \citep{levin2020,manzorro2022}, and propose to process one atomic column at a time. The algorithm focuses on the cropped videos which spatially correspond to the approximate regions of atomic columns, and outputs one single ridge curve after every run.
Given the data's discrete nature, diverse trajectory curvatures and the potential degeneration, our investigation with anticipated continuous output is challenging and novel.

This paper is organized as follows. \Cref{sec:pre} introduces preliminaries including notations, necessary setup assumptions, and preprocessing steps. The details of our methodology are introduced in \autoref{sec:method}.
% Corresponding to the ridge properties summarized in \Cref{sec:prop}, \Cref{sec:quant} proposes the measures to perform the detection job, while \Cref{sec:curve} connects the detected points to form the continuous functional trajectory.
\Cref{sec:uncertainty} delivers some elementary and discrete results for uncertainty summaries about an estimated curve. Supported by simulated results from \Cref{sec:simu}, \Cref{sec:app} demonstrates the superior performance of our method in our motivating TEM application (above) in material science. \Cref{sec:con} concludes with discussion.
% Some background introductions on the physics side are included in \autoref{app:syn}.
\autoref{app:grad} introduces an optional alternative schema of methodology development that supplements and partially updates \autoref{sec:method}. Proofs for selected propositions from throughout the paper appear in \autoref{app:proof}.

\section{Preliminaries}\label{sec:pre}

Throughout this paper, a video is described by a sequence (length $T$) of gray-scale images, each with size $M \times N$. The set of pixel indices is given by the collection $\Omega = \{(m, n, \tau): m \in [M]; n \in [N]; \tau \in [T]\}$ where we introduce the notation $[k] = \{0, 1, \dots, k-1\}$ for integer $k$. In addition, for $\tau \in [T]$, we denote the $t = \tau$ image frame as $\Omega(\tau) = \Omega \bigcap \{(x,y,t): t = \tau\}$. 

The pixel values of the image sequence are discretely evaluated by a mapping $f(x,y,t): \mathbb{R}^3 \to \mathbb{R}$ at the triplet grid locations $(x,y,t) = (m, n, \tau) \in \Omega$, denoted as $f_{m,n,\tau}$. A continuous ridge curve, which may not be restricted to the lattice $\Omega$, is defined as $\gamma: \mathbb{R} \to \mathbb{R}^3$. 

Denote $e_t$ as the indicator vector for the temporal dimension, i.e., $e_t = (0, 0, 1)$. For a matrix $A$, $\trace(A)$ represents the trace and $A^+$ denotes its Moore-Penrose general inverse. The operator $x_+$ computes $\max(x, 0)$ for any scalar $x$. For two vectors $a$ and $b$ of the same dimension, we denote $\big<a, b\big>$ as their dot product, and $\cos(a, b) = \frac{\langle a, b \rangle}{\|a\| \cdot \|b\|}$ as the cosine similarity between them. The indicator function $\mathbbm{1}(X)$ takes value 1 (or 0) if the statement $X$ is true (or false).

We define a ridge $\gamma$ as a continuous collection of points on a three-dimensional mapping $f$ which graphically resembles the tube-shaped object as demonstrated in \autoref{fig:blob_ridge}. Here the definition gets slightly extended from some other similar alternatives (e.g., \citealt{porteous2001}), see \autoref{enum:Quad} below, to allow for degenerate eigenvalues. In consequence, the trajectory curve in our TEM application will not get suspended at those extraordinary image frames when the atomic column is absent. \\

\begin{defn}\label{def:ridge}
Given a second-order differentiable mapping $f: \mathbb{R}^3 \to \mathbb{R}$, denote $\nabla_f(p) \in \mathbb{R}^3$ and $\Delta_f(p) \in \mathbb{R}^{3 \times 3}$ as the gradient vector and the Hessian matrix of the mapping $f$ at point $p \in \mathbb{R}^3$, respectively. Additionally, assume the Hessian matrix $\Delta_f(p)$ has eigenvalues $\lambda_{f,1}(p), \lambda_{f,2}(p), \lambda_{f,3}(p)$ and corresponding unit eigenvectors $v_{f,1}(p), v_{f,2}(p), v_{f,3}(p) \in \mathbb{R}^3$. Then, $p$ is a point on a ridge of $f$, denoted by $\gamma \subset \mathbb{R}^3$, if
\begin{enumerate}
    \item \label{enum:Tang} $\big< \nabla_f(p), v_{f,2}(p) \big> = \big< \nabla_f(p), v_{f,3}(p) \big> = 0$, i.e., the gradient $\nabla_f(p)$ is parallel to the Hessian eigenvector $v_{f,1}(p)$. Consequently, the tangent direction of the ridge $\gamma$ at $p$ is characterized by $v_{f,1}(p)$; % and $\Delta_f^+(p) \nabla_f(p) \parallel v_{f,1}(p)$;
    \item \label{enum:Quad} $0 \geq \lambda_{f,2}(p) \geq \lambda_{f,3}(p)$;
    \item \label{enum:Magn} Furthermore, the curvature of the mapping $f$ along direction $v_{f,1}(p)$ is small or relatively small, i.e., $|\lambda_{f,i}(p)| \gg |\lambda_{f,1}(p)|$ or $\left|\frac{\lambda_{f,i}(p)}{\lambda_{f,1}(p)}\right| \gg 1$ for both $i = 2$ and $3$, where $\gg$ denotes much greater than.
\end{enumerate}
\end{defn}

\begin{figure}[htbp]
    \centering
    \includegraphics[width=\textwidth]{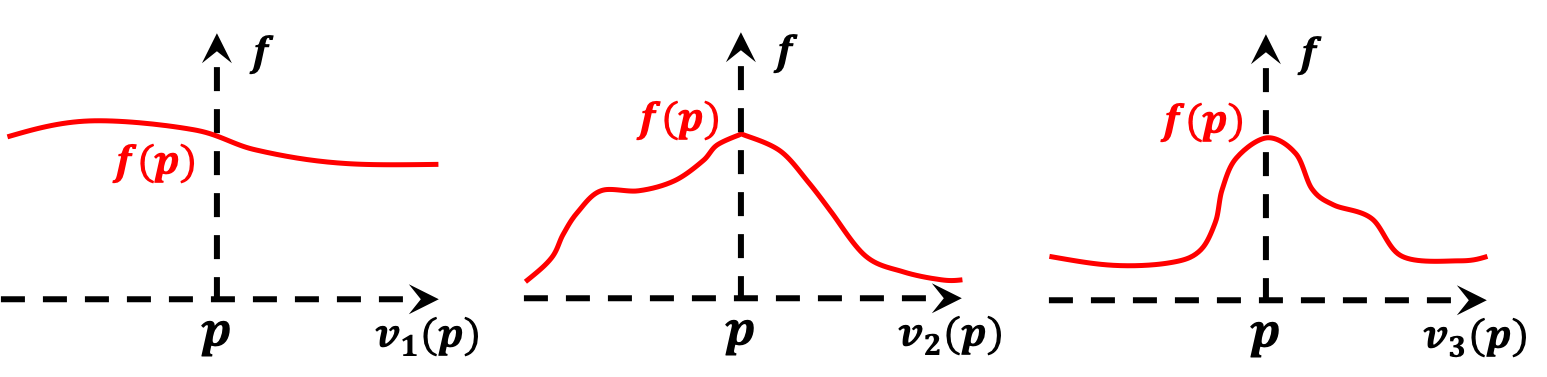}
    \caption{The projected behaviors of the mapping $f$ at a ridge point $p \in \gamma$ along the axes of a local coordinate system. Here the basis of the new coordinate system is formed by the set of orthonormal Hessian eigenvectors $\{v_1(p), v_2(p), v_3(p)\}$.}
    \label{fig:cross_section}
\end{figure}

\begin{rem}\label{rem:def}
\begin{enumerate}[(a)]
    \item The \autoref{fig:cross_section} graphically illustrates \autoref{def:ridge}. Generally speaking, if we introduce at point $p$ a local coordinate system different from the classical Cartesian $(x,y,t)$ representations, and specify the axes using the local Hessian eigenvectors $\{v_{f,1}(p), v_{f,2}(p), v_{f,3}(p)\}$, then $p$ is on a ridge if the mapping $f$ attains the local maximum at $p$ along two of the three axes directions; see the latter two subplots in \autoref{fig:cross_section}. In the TEM video application, these two axes form the plane that is usually close to the image frame.

    \item The definition of valleys only differs from \Cref{def:ridge} in \autoref{enum:Quad} with all the inequality signs flipped, i.e., a ridge point $p$ of the mapping $f$ is naturally a valley point of $-f$. With the shift- and scale-free properties imposed through later development, our method can directly take the negated TEM video as the input for implementation.

    \item In our later analysis, especially when concerning the TEM application, the video mapping $f$ is characterized by the discrete pixel values and remains fixed. We will omit the subscript $f$ for conciseness.
\end{enumerate}
\end{rem}

%Though introduced for the ridge case, the method of this paper has broader use cases and can be immediately and equivalently applied to valleys.

We then impose an essential setup assumption to fix the temporal axis as the major stretch direction of the ridge $\gamma$.

\begin{ass}\label{ass:temporal}
There exists one unique continuous ridge curve $\gamma$ within the video region $[0,M) \times [0,N) \times [0,T)$ that can be parameterized temporally, i.e., $\forall t \in [0,T)$, there exists one unique ridge mapping
\begin{equation}\label{eq:curve}
    \gamma(t) = (u(t), w(t), t) \in [0, M) \times [0, N) \times [0, T),
\end{equation}
where $u(t)$ and $w(t)$ are corresponding spatial coordinates functions. Furthermore, the ridge has tangent $\gamma'(t) = (u'(t), w'(t), 1)$.
\end{ass}

\begin{rem}
     In \autoref{ass:temporal}, we impose the existence and uniqueness restrictions, as they comply with our proposed TEM application which processes one atomic column at a time given the cropped TEM videos. The parameterization \eqref{eq:curve} is motivated by the application on TEM videos, where like \autoref{fig:blob_ridge}, the spheres of the atomic columns from the TEM video are stacked relatively stably across time to form a tube. In addition, \eqref{eq:curve} is beneficiary for follow-up analysis. For example, given the input of any frame index $\tau \in [T]$, the ridge function $\gamma$ can then directly return the corresponding spatial location of the selected atomic column within the specific frame.
\end{rem}

The computation of first- and second-order derivatives are essential in \Cref{def:ridge}, especially with the discrete nature of the pixel lattice $\Omega$. We propose to convolve with Gaussian functions not only to smooth the noisy data but also to give better behaved differentials.

\begin{defn}\label{def:discrete}
Denote the three-dimensional Gaussian function% kernel
$$
G(x, y, t; \sigma, \delta) = \frac{1}{2 \pi \sigma^2 \cdot \sqrt{2 \pi \delta^2}} \exp(-\frac{x^2 + y^2}{2 \sigma^2} - \frac{t^2}{2 \delta^2}),
$$
where $\sigma$ and $\delta$ represent spatial and temporal scales, respectively.
Then, given $p \in \Omega$, the scaled gradient vector and Hessian matrix
\begin{equation}\label{eq:gradient}
    \widetilde{\nabla}(p) = (\sigma\widetilde{\nabla}_{x}(p), \sigma\widetilde{\nabla}_{y}(p), \delta\widetilde{\nabla}_{t}(p))' \quad \mbox{\&} \quad \widetilde{\Delta}(p) = \begin{pmatrix}
        \sigma^2 \widetilde{\Delta}_{xx}(p) & \sigma^2 \widetilde{\Delta}_{xy}(p) & \sigma\delta \widetilde{\Delta}_{xt}(p) \\
        \sigma^2 \widetilde{\Delta}_{yx}(p) & \sigma^2 \widetilde{\Delta}_{yy}(p) & \sigma\delta \widetilde{\Delta}_{yt}(p) \\
        \sigma\delta \widetilde{\Delta}_{tx}(p) & \sigma\delta \widetilde{\Delta}_{ty}(p) & \delta^2 \widetilde{\Delta}_{tt}(p)
    \end{pmatrix}
\end{equation}
of the mapping $f$ are approximated by evaluating the elements of partial derivatives via discrete convolutions analogous to
$$
\begin{aligned}
\widetilde{\nabla}_{x}(p) & = \frac{\partial}{\partial x} \Big( \sum_{(m, n, \tau) \in \Omega} f_{m,n,\tau} G(x-m, y-n, t-\tau; \sigma, \delta) \Big)_{(x,y,t) = p},\\
\widetilde{\Delta}_{xx}(p) & = \frac{\partial^2}{\partial x^2} \Big( \sum_{(m, n, \tau) \in \Omega} f_{m,n,\tau} G(x-m, y-n, t-\tau; \sigma, \delta) \Big)_{(x,y,t) = p}.
\end{aligned}
$$
\end{defn}

\begin{rem}
    Though both multiplied in the scale-space derivatives \eqref{eq:gradient} \citep{lindeberg1998}, the scale $\delta$ along the temporal axis merely serves as the denoising parameter through convolution, while its spatial counterpart $\sigma$ plays its additional role for automatic scale adaptation \citep{lindeberg1996}. In practice, the spatial scale $\sigma$ usually implies the size of the atomic column in the TEM images, and is tuned with references from the cross-sectional radii of the tube-shaped object as in \autoref{fig:blob_ridge}. On the other hand, the temporal scale $\delta$ is often set small to preserve as much subtle dynamics as possible to accurately recover the ridge's local curvatures, which correspond to the minor movements of the interested atomic column in the TEM application.
\end{rem}

% We then impose the following assumption to avoid singular derivatives in computation.

\begin{ass}\label{ass:nonzero}
The gradient vectors $\widetilde{\nabla}(p)$ are non-singular.
\end{ass}

\begin{rem}
     \autoref{ass:nonzero} is imposed to avoid singular derivatives in computation. It is not restrictive, given the noise of the image dataset and the convolution applied in preprocessing. It only restricts the estimators with noise present hence does not rule out the possibilities of those underlying singular cases that we aim to tackle.
\end{rem}

The next definition standardizes and updates the estimated Hessian matrix $\widetilde{\Delta}(p)$ in order to improve the universal applicability of our method.

\begin{defn}\label{def:standard}
    Set $\mu$ as the median of the gradients norms $\{\|\widetilde{\nabla}(p)\|: p \in \Omega\}$. The Hessian matrices are updated by
    \begin{equation}\label{eq:updateHess}
        \widetilde{\Delta}(p) \mapsto \widetilde{\Delta}(p) / \mu.
    \end{equation}
\end{defn}

\begin{rem}\label{rem:standard}
    \begin{enumerate}[(a)]
        \item The update \eqref{eq:updateHess} for the Hessian matrices is introduced to ensure the scale-free property of our approach, so that the developed algorithm can be universally implemented on various image sequences with different levels of pixel intensities.

        \item For simplicity hereafter, the Hessian $\Delta(p)$ and its approximation $\widetilde{\Delta}(p)$ in \eqref{eq:updateHess} will share the same notations of eigenvalues $\lambda_{i}(p)$ and eigenvectors $v_{i}(p)$ except where otherwise stated.
    \end{enumerate}
\end{rem}

\section{Methodology}\label{sec:method}

In \autoref{sec:prop}, we first explore some analytical properties of the points on the ridge in the geometrical sense. Then \autoref{sec:quant} quantifies every grid point's fulfillment of these properties, and proposes a ridge score accordingly. Finally, \autoref{sec:curve} introduces the proposed non-parametric algorithm for the continuous ridge curve.

In addition, an optional but recommended supplementary schema is discussed in \autoref{app:grad}. Under specific scenarios to be elaborated therein, it potentially improves the power of the algorithm.

Note that for our later development, we stick to \autoref{ass:temporal} that one and only one ridge curve will be estimated from the input video. In terms of the TEM application, it means that the algorithm will be implemented on manually cropped videos to extract the trajectories of atomic columns one at a time.

\subsection{Properties}\label{sec:prop}

We start with \autoref{prop:approx} summarising some properties of the points along the ridge. The proof of it can be found in \autoref{app:proof}.

\begin{prop}\label{prop:approx}
If the ridge $\gamma$ of the mapping $f$ passes near a lattice point $p \in \Omega$, then
\begin{enumerate}
    \item According to \autoref{enum:Tang} of \autoref{def:ridge}, the ridge's local direction is approximated by the eigenvector of the estimated Hessian $\widetilde{\Delta}(p)$:
    \begin{equation}\label{eq:v}
        v(p) = v_{1}(p).
    \end{equation}
    In addition, the vector $v(p)$ is redirected such that $\big< v(p), e_t \big> \ge 0$.

    \item The proxy of the ridge's local curvature is given by the eigenvalue of the estimated Hessian $\widetilde{\Delta}(p)$:
    \begin{equation}\label{eq:lambda}
        \lambda(p) = \lambda_{1}(p).
    \end{equation}
    
    \item The curve's spatial perturbation level can be measured by the cosine similarities between the ridge's local direction \eqref{eq:v} and the temporal indicator $e_t$:
    \begin{equation}\label{eq:rho}
        \rho(p) = \big| \cos\big( v(p), e_t \big) \big|.
    \end{equation}

    \item The satisfaction of \autoref{enum:Tang} in \Cref{def:ridge} can be quantified by the cosine similarities which measure the angular difference between the ridge's local direction \eqref{eq:v} and the mapping's underlying landscape gradient:
    \begin{equation}\label{eq:theta}
        \theta(p) = \Big| \cos\big( v(p), \widetilde{\nabla}(p) \big) \Big|.
    \end{equation}
    
    \item Concerning \Cref{enum:Quad,,enum:Magn} of \Cref{def:ridge}, the following quantities approximately summarize some behaviors of the Hessian eigenvalues:
    \begin{equation}\label{eq:kappa}
    \begin{dcases}
        \eta(p) = \frac{2 \lambda_{2}(p) \lambda_{3}(p)}{\lambda^2_{2}(p) + \lambda^2_{3}(p)}\\
        \kappa(p) = \lambda_{2}(p) \lambda_{3}(p) - \lambda^2(p)
    \end{dcases}.
    \end{equation}
\end{enumerate}
\end{prop}

In \autoref{app:grad}, \autoref{cor:approx} proposes parallel approximations to the above quantities denoted in \autoref{prop:approx}. The approximations bring potential benefits in both theoretical interpretations and practical performances under specific conditions. Refer to \autoref{app:grad} for more details.

The following property and its successive remark
% , with major focus on individual pixels, 
are summarised based on \autoref{prop:approx} as the initial reference criteria to filter the ridge trajectory in a temporally stacked image series.

\begin{property}[Intra-Frame]\label{property:ridge}
The grid point $p \in \Omega$ is likely to be on the ridge $\gamma$ if:
\begin{enumerate}
    \item Either of the following conditions hold for the ridge's local direction $v(p)$:
    \begin{enumerate}[a.]
        \item \label{enum:temporal} $v(p)$ is maximally parallel to the temporal indicator $e_t$, i.e., $\rho(p) \approx 1$.
        \item \label{enum:theta} $v(p)$ optimally approximates the mapping's gradient, i.e., $\theta(p) \approx 1$.
    \end{enumerate}
    \item \label{enum:concave} The two Hessian eigenvalues $\lambda_2(p)$ and $\lambda_3(p)$ have the same sign as well as similar magnitudes, i.e., $\eta(p) \approx 1$. In addition, they are relatively larger in absolute values compared to the curvature proxy $\lambda(p)$ along the ridge, hence $\kappa(p)$ is positive and relatively large in magnitude.
\end{enumerate}
\end{property}

\begin{rem}
    \begin{enumerate}[(a)]
        \item The \autoref{enum:temporal} of \autoref{property:ridge} is based on our understanding that a smooth ridge curve often times has relatively spatial-stable tangent directions that likely lie on the temporal axis. In other words, the ridge curve, or the location of the atomic column, stays almost put spatially. It is indeed highly probable near the degeneration scenarios in the TEM applications, as one may expect the absent atomic column to reappear after a period of time at the same place where it disappeared.

        \item The \autoref{enum:theta} of \autoref{property:ridge} is summarized from \eqref{eq:theta} and its relevant claims in \autoref{prop:approx}. Note that \Cref{enum:temporal,enum:theta} provide criteria from two different perspectives and can both be nearly optimal at most near-ridge pixels. Meanwhile, they have detection preferences such that their exact optimums may not be attained simultaneously. For instance:\begin{enumerate}[(i)]
            \item \autoref{enum:temporal} dominates when a nearly singular gradient is encountered, i.e., $\| \widehat{\nabla}(p) \| \approx 0$. In the TEM application, it most likely happens when the atomic column is absent;
            \item \autoref{enum:theta} dominates when the ridge has apparent spatial movement. In the TEM application, it happens when the atomic column is drifting.
        \end{enumerate}

        \item The similar magnitudes of $\lambda_{2}(p)$ and $\lambda_{3}(p)$, as in the \autoref{enum:concave} of \autoref{property:ridge}, result in a nearly spherical ellipse on the ridge tube's cross section, with the two symmetric axes characterized by the corresponding eigenvectors. Such cross-sectional elliptical pattern is usually similar to the shape of an atomic column in a TEM image.

        \item Overall, the \autoref{property:ridge} focuses on the local gradient and Hessian behaviors of individual pixels restricted within a single frame, hence is tagged as \emph{Intra-Frame}.
    \end{enumerate}
\end{rem}

While \Cref{property:ridge} is better at depicting the behaviors of the pure ridge pixels, we expect our work to be more effective for the generalizations or the occasionally absent atomic columns in the TEM application. Indeed, in addition to \autoref{enum:temporal}, the applicability under these extreme cases can be further addressed by explicitly enforcing the continuity constraint along the curve. In practice, we exploit the direction vectors $v(p)$ in \eqref{eq:v}, and penalize on the functional second-order roughness along the curve \citep{green1994}.

Prior to penalization, we first introduce the following definition of the candidate ridge tangent.

\begin{defn}\label{def:tangent}
For $p \in \Omega$, define the candidate ridge tangent estimator
\begin{equation}\label{eq:candidate}
    v_\gamma(p) = \frac{v(p)}{\big< v(p), e_t \big>},
\end{equation}
where we denote its coordinates as $v_\gamma(p) = (u'_\gamma(p), w'_\gamma(p), 1)$.
\end{defn}

\begin{rem}
    \begin{enumerate}[(a)]
        % \item Note that the \autoref{def:tangent} involves an enhanced directional approximation $\bar{v}(p)$. It will be defined later as \eqref{eq:vbar} in \Cref{subsec:outstand}, which involves both $v(p)$ and $\hat{v}(p)$.
        
        \item We name the vector $v_\gamma(p)$ in \eqref{eq:candidate} as the candidate ridge tangent because it is treated as the ridge tangent that points outwards from arbitrary $p \in \Omega$ as if the pixel $p$ were on the ridge curve $\gamma$, given \Cref{def:ridge} and \Cref{ass:temporal}. Concerning the TEM application, if the interested atomic column were located at $p$ from the $\tau$-th frame $\Omega(\tau)$, then the spatial components of $v_\gamma(p)$ indicate the selected atomic column's potential movement direction at that moment.
        
        \item The transformation \eqref{eq:candidate} rescales the last element of the candidate ridge tangent to be 1, so that $v_\gamma(p)$ matches with $\gamma'(t)$ in \autoref{ass:temporal} regarding the temporal element.

        % \item The estimator $v_\gamma(p)$ partially inherits the strengths of the gradient-yielded approximation $\hat{v}(p)$ as mentioned in \autoref{rem:features}, especially when the pixels are within the neighborhood of non-degenerated ridge segments. In particular, it is likely to point towards the ridge curve due to the mapping's gradient landscape.
    \end{enumerate}
\end{rem}

Below we demonstrate the second property statement which supports the idea of penalizing the curve's roughness.

\begin{property}[Inter-Frame]\label{property:cont}
Given the frame index $\tau \in [T]$, if the grid points $p = (m, n, \tau) \in \Omega(\tau)$ and $p^* = (m^*, n^*, \tau^*) \in \Omega(\tau)$, $\tau^* = \tau + 1$, are two temporally sequential grid points that are close to the ridge curve $\gamma$, and have their candidate tangents $v_\gamma(p)$ and $v_\gamma(p^*)$ derived from \Cref{def:tangent} respectively, then the following conditions hold:
\begin{enumerate}
    \item The ridge passes near $p$ with almost parallel tangent, i.e., $\gamma(\tau) \approx p$, $\gamma'(\tau) \approx v_\gamma(p)$; and analogously for $p^*$, i.e., $\gamma(\tau^*) \approx p^*$, $\gamma'(\tau^*) \approx v_\gamma(p^*)$;
    \item Denote the functional second-order roughness of the ridge as the integral
    $$
    \int_{\tau}^{\tau^*} \|\gamma''(t)\|^2 dt = \int_{\tau}^{\tau^*} \big( |u''(t)|^2 + |w''(t)|^2 \big) dt,
    $$
    and it is small in magnitude.% as its smaller value implies better smoothness within the curve segment.
\end{enumerate}
\end{property}

\begin{rem}
    Note that the smaller roughness value implies better smoothness within the curve segment. The tag \emph{Inter-Frame} is given due to the fact that \autoref{property:cont} focuses on the interactive continuity restraints between pixels from consecutive frames. Under the setting of the TEM application, the design of the roughness aims to penalize those extremely volatile movements of the selected atomic column.
\end{rem}

\subsection{Ridge Quantification}\label{sec:quant}
This section introduces measures based on \Cref{property:ridge,property:cont} to obtain the likeliness that whether a grid point $p \in \Omega$ is on the ridge $\gamma$. Particularly, \Cref{subsec:outstand} designs an intra-frame weight that aggregates metrics from \Cref{property:ridge} with pixel-wise gradient and Hessian information, while \Cref{subsec:continuity} proposes the final inter-frame weight involving the roughness penalization according to \Cref{property:cont}.
% \Cref{subsec:summary} aggregates the measures and summarises the section.

\subsubsection{Intra-Frame Pixel-Wise Standout Measures}\label{subsec:outstand}

To understand how one pixel stands out within an image frame giving it higher weight to be a point assigned to the ridge curve, we define the following terms as the intra-frame metrics to be the immediate quantification of \autoref{property:ridge}.

\begin{defn}[Intra-Frame Metrics]\label{def:intra}
For $p \in \Omega$, the following measures are defined to account for:\begin{enumerate}
    \item Spatial stability (\autoref{enum:temporal} of \autoref{property:ridge}):
    \begin{equation}\label{eq:Lrho}
        L_\rho(p) = 2 \rho(p);
    \end{equation}

    \item First-order extremum (\autoref{enum:theta} of \autoref{property:ridge}):
    \begin{equation}\label{eq:Ltheta}
        L_\theta(p) = 2 \theta(p);
    \end{equation}

    \item Second-order concavity (\autoref{enum:concave} of \autoref{property:ridge}):
    \begin{equation}\label{eq:Letakappa}
        L_{\eta, \kappa}(p) = 2 \eta(p) + 2 \eta(p) \log(1 + \kappa_+(p)).
    \end{equation}
\end{enumerate}
\end{defn}

Given our constructions, these metrics can be shown to be effective for filtering out the ridge points from pixel grids. In particular, supplementary to \Cref{def:intra}, the following proposition summarizes these metrics' behaviors and provides evidence for these choices. In short, the grid point $p$ with higher metric values are more likely to be on the ridge, and such statement is quantitatively consistent with \autoref{property:ridge}. The corresponding proof is in \autoref{app:proof}.

\begin{prop}\label{prop:intra_behavior}
    The metrics from \autoref{def:intra} satisfy the following statements:\begin{enumerate}
        \item The less the angular difference between the ridge's local direction $v(p)$ and the temporal indicator $e_t$, the larger the metric value for $L_\rho(p)$ \eqref{eq:Lrho}. In addition, $L_\rho(p)$ attains its maximum if and only if the angular difference is zero.

        \item The less the angular difference between the ridge's local direction $v(p)$ and the gradient estimator $\widetilde{\nabla}(p)$, the larger the metric value for $L_\theta(p)$ \eqref{eq:Ltheta}. In addition, $L_\theta(p)$ attains its maximum if and only if the angular difference is zero, i.e., \autoref{enum:Tang} of \Cref{def:ridge} is satisfied approximately.

        \item The metric $L_{\eta, \kappa}(p)$ \eqref{eq:Letakappa} favors the pairs of eigenvalues $\big( \lambda_2(p), \lambda_3(p) \big)$ with same signs and similar magnitudes, as well as larger magnitudes compared to $\lambda_1(p)$.
    \end{enumerate}
\end{prop}

The following definition then combines the intra-frame metrics from \autoref{def:intra} to the pixel-wise weights, and quantitatively summarizes \autoref{prop:intra_behavior}.

\begin{defn}\label{def:weight}
Given a frame $\Omega(\tau)$ for $\tau \in [T]$, the local measures in \Cref{def:intra} collectively give the initial ridge weights for $p \in \Omega(\tau)$
\begin{equation*}\label{eq:weight}
    \Phi(p) = \exp\big( L_\rho(p) + L_\theta(p) + L_{\eta,\kappa}(p) \big).
\end{equation*}
\end{defn}

In \autoref{app:grad}, \autoref{cor:weight} is presented as an optional update to \autoref{def:weight} utilizing parallel approximations from \autoref{cor:approx}, with potential algorithmic enhancements under specific conditions; refer to \autoref{app:grad} for more details.

% \begin{rem}
%     The \autoref{def:weight} merges the two versions of estimations and metric designs (hatted and unhatted, notation-wise speaking). It combines the featured strength of the two systems listed in \autoref{rem:features} for enhanced estimation. In particular, the (unhatted) terms $\varphi(p)$ and $v(p)$ will supposedly dominate in \eqref{eq:weightFinal} and \eqref{eq:vbar} under degeneration scenarios, while the hatted ones will play more essential roles within a neighborhood of non-degenerated segments.
% \end{rem}

\subsubsection{Inter-Frame Continuity Penalization}\label{subsec:continuity}
In addition to considering the intra-frame quantification, our \autoref{property:cont} advocates the curve smoothness especially to enhance the compatibility in the TEM application. We have the following metric definition that quantifies the continuity between pixels from consecutive frames.

\begin{defn}[Inter-Frame Metric]\label{def:rough}
Given $\tau \in [T]$, consider the two temporally consecutive points $p \in \Omega(\tau)$ and $p^* \in \Omega(\tau^*)$ where $\tau^* = \tau + 1$. Assume a cubic functions $\zeta_{p,p^*}(t) = (u_{p,p^*}(t), w_{p,p^*}(t))$ locally interpolates the spatial coordinates of the two pixels, then define the roughness metric of the pixel pair
\begin{equation}\label{eq:continuity}
    \psi(p,p^*) = \exp \Big( - \frac{1}{2} \int_{\tau}^{\tau^*} \| \zeta''_{p,p^*} (t) \|^2 dt \Big),
\end{equation}
where $\| \zeta''_{p,p^*} (t) \|^2 = \big( u''_{p,p^*}(t) \big)^2 + \big( w''_{p,p^*}(t) \big)^2$.
\end{defn}

\begin{figure}[htbp]
    \captionbox{Illustration of \autoref{def:rough} for the interpolation and the roughness penalization.\label{fig:interpolate}}[.45\textwidth]
        {\includegraphics[width=.45\textwidth]{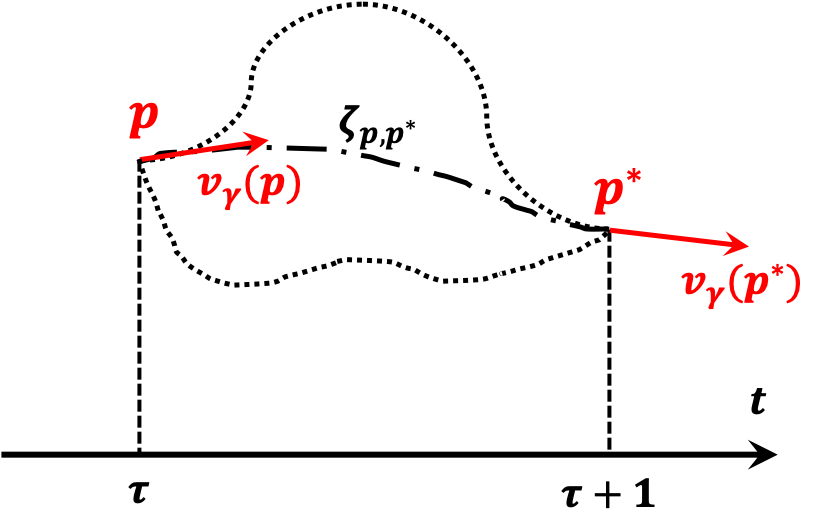}}\hspace{.1\textwidth}%
    \captionbox{Illustration of \autoref{def:score} for the forward and backward metrics.\label{fig:markov}}[.35\textwidth]
        {\includegraphics[width=.35\textwidth]{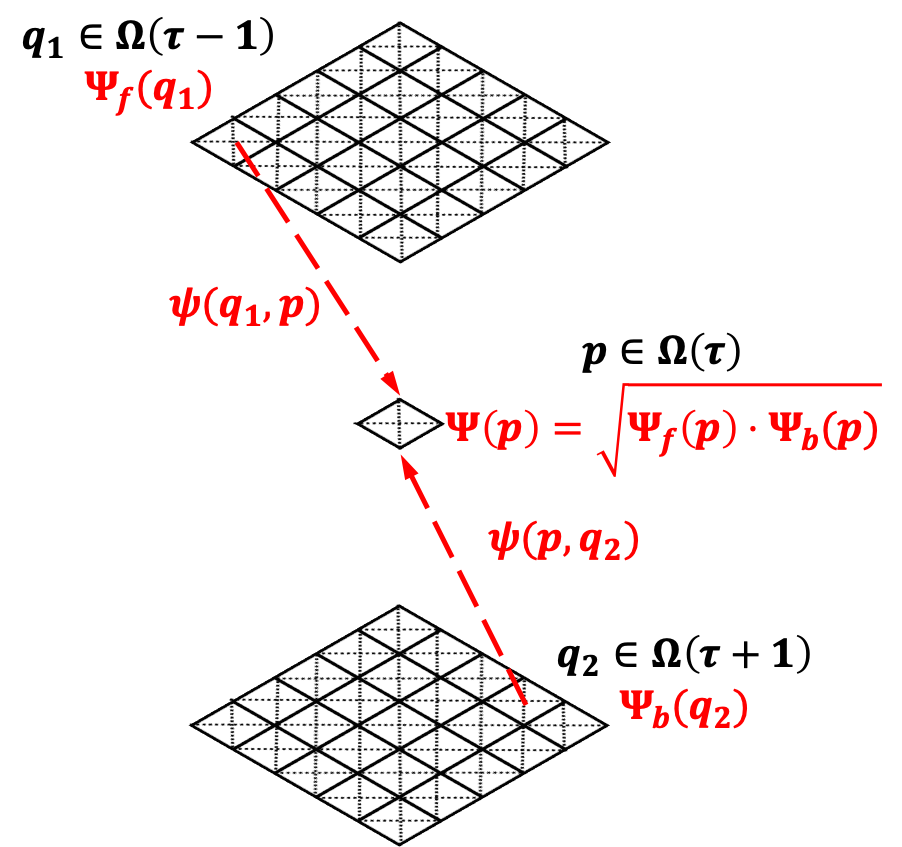}}
\end{figure}

\begin{rem}
    \begin{enumerate}[(a)]
        \item The \autoref{fig:interpolate} illustrates the interpolation idea in \autoref{def:rough}. To summarize, the cubic function $\zeta_{p, p^*}$ not only connects the two pixels $p$ and $p^*$, but also has the tangent directions that coincide with the spatial components of their candidate ridge tangent $v_\gamma(p) = (u_\gamma'(p), w_\gamma'(p), 1)$ and $v_\gamma(p^*) = (u_\gamma'(p^*), w_\gamma'(p^*), 1)$ at $p$ and $p^*$, respectively. The candidate ridge tangents are transformed from the local directions $v(p)$ according to \autoref{def:tangent}.

        \item In practice, assume $p = (m, n, \tau)$, $p^* = (m^*, n^*, \tau^*)$, and $\zeta_{p,p^*}(t) = (u_{p,p^*}(t), w_{p,p^*}(t))$ where
        \begin{align*}
            u_{p, p^*}(t) = a_{p,p^*}^u t^3 + b_{p,p^*}^u t^2 + c_{p,p^*}^u t + d_{p,p^*}^u, \qquad w_{p, p^*}(t) = a_{p,p^*}^w t^3 + b_{p,p^*}^w t^2 + c_{p,p^*}^w t + d_{p,p^*}^w.
        \end{align*}
        Without loss of generality, set $\tau = 0$ and $\tau^* = 1$, then
        \begin{itemize}
            \item $\zeta_{p,p^*}(0) = (d_{p,p^*}^u, d_{p,p^*}^w) = (m, n)$;
            \item $\zeta_{p,p^*}(1) = (a_{p,p^*}^u + b_{p,p^*}^u + c_{p,p^*}^u + d_{p,p^*}^u, a_{p,p^*}^w + b_{p,p^*}^w + c_{p,p^*}^w + d_{p,p^*}^w) = (m^*, n^*)$;
            \item $\zeta_{p,p^*}'(0) = (c_{p,p^*}^u, c_{p,p^*}^w) = \big( u'_\gamma(p), w'_\gamma(p) \big)$;
            \item $\zeta_{p,p^*}'(1) = (3 a_{p,p^*}^u + 2b_{p,p^*}^u + c_{p,p^*}^u, 3 a_{p,p^*}^w + 2 b_{p,p^*}^w + c_{p,p^*}^w) = \big( u'_\gamma(p^*), w'_\gamma(p^*) \big)$.
        \end{itemize}
        The coefficients can be solved from the linear system.
        
        \item By integrating the squared norm of the second order derivative, the metric \eqref{eq:continuity} measures the perturbation level of the interpolation $\zeta_{p,p^*}$. Specifically, higher metric values imply more smoothly connected local functional segments, and consequently less abrupt spatial movements of the selected atomic column concerning the TEM application.
    \end{enumerate}
\end{rem}

The subsequent definition then incorporates the pixel-wise weight from \autoref{def:weight} and finalizes the ridge score with the smoothness penalization from \autoref{def:rough}; see \autoref{fig:markov} for the illustration.

\begin{defn}\label{def:score}
Given $p \in \Omega(\tau)$, $\tau \in [T]$, we recursively define the forward and backward accumulated metrics $\big( \Psi_f(p)$, $\Psi_b(p) \big)$ as
$$
\Psi_f(p) = \Phi(p) \cdot \big( \sum_{q_1 \in \Omega(\tau-1)} \Psi_f(q_1) \psi(q_1,p) \big), \hspace{1cm} \Psi_b(p) = \Phi(p) \cdot \big( \sum_{q_2 \in \Omega(\tau+1)} \Psi_b(q_2) \psi(p,q_2) \big).
$$
Then the geometric mean yields the final metric
\begin{equation}\label{eq:measure}
\Psi(p) ~ \propto ~ \sqrt{\Psi_f(p) \cdot \Psi_b(p)}.
\end{equation}
Furthermore, the metrics are normalized to satisfy $\sum_{p \in \Omega(\tau)} \Psi(p) = 1.$
\end{defn}

\begin{rem}
    When designing the forward (backward) metrics above, we regard the ridge's trajectory along the (reversed) sequence of image frames as a Markov process. For instance, the ridge curve (or the atomic column's trajectory) may reach the pixel $p \in \Omega(\tau)$ potentially from any pixel $q_1$ in the previous frame $\Omega(\tau-1)$ by a forward transition. The forward metric $\Psi_f(p)$ is hence cumulatively calculated by summing up the probabilities of all these possible forward transitions that originate from $\Omega(\tau-1)$. The roughness metric \eqref{eq:continuity} in \autoref{def:rough} is utilized as the transition probability between the two pixels. Compared to the ordinary arithmetic mean, the geometric mean \eqref{eq:measure} can better downplay the weight of those pixels whose forward and backward accumulative metrics have inconsistent behaviors.
\end{rem}

\subsection{Non-parametric Curve Connection}\label{sec:curve}
As \autoref{ass:temporal} suggests, the curve trajectory is parameterized as follows:
\begin{equation*}
    \gamma(t) = (u(t), w(t), t) \in \mathbb{R}^3.
\end{equation*}
To connect the ridge curve non-parametrically, we consider the pixels within every image frame $\Omega(\tau)$ as an ensemble, and proceed to the ridge estimation with ensemble summaries. In particular, for each frame $\Omega(\tau)$, the pixels $p \in \Omega(\tau)$ and their corresponding direction vectors $v(p)$ yield an aggregated estimator of the frame-specific functional element. The following definition and remark detail the implementation and its intuitive interpretations.
% To connect the ridge curve non-parametrically, one may proceed with either of the two frameworks:\begin{enumerate}
%     \item Pixel-wise: every pixel $p \in \Omega$ and its candidate direction $v(p)$ corresponds to a functional element, and contribute to the ridge with corresponding weight $\Psi(p)$.
    
%     \item Frame-wise: within temporal frame $\Omega(\tau)$, every pixel $p$ with corresponding candidate direction $v(p)$ and weight $\Psi(p)$ yield an aggregated estimator of functional element that contributes to the recovery of the ridge curve.
% \end{enumerate}
% In this project, the latter one is elaborated below due to its intuitive interpretations and straightforward implementations.

\begin{defn}\label{def:nonpar}
Within a temporal frame $\Omega(\tau)$ for $\tau \in [T]$, define the weighted averages
\begin{equation}\label{eq:loc}
    \bar{p}(\tau) = \sum_{p \in \Omega(\tau)} p \Psi(p), \qquad \bar{v}_\gamma(\tau) = (\bar{u}(\tau), \bar{w}(\tau), 1) ~ \propto \sum_{p \in \Omega(\tau)} v(p) \Psi(p),
\end{equation}
and the frame-local linear functional element
\begin{equation}\label{eq:loc_lin}
    \bar{\gamma}_\tau(t) = \bar{p}(\tau) + \bar{v}_\gamma(\tau) \cdot  (t - \tau).
\end{equation}
Then given kernel function $K(t)$ and bandwidth $h$, the curve is non-parametrically estimated
\begin{equation}\label{eq:nonpar}
    \bar{\gamma}(t) = \frac{\sum_{\tau \in [T]} \bar{\gamma}_\tau(t) \cdot K\Big(\frac{t - \tau}{h}\Big)}{\sum_{\tau \in [T]} K\Big(\frac{t - \tau}{h}\Big)}.
\end{equation}
\end{defn}

\begin{rem}
    \begin{enumerate}[(a)]
        \item For the weighted averages in \eqref{eq:loc}:\begin{enumerate}[(i)]
            \item $\bar{p}(\tau)$ is considered as the candidate intersection between the image frame $\Omega(\tau)$ and the ridge $\gamma$, which under the TEM setting corresponds to the frame-aggregated location estimator of the selected atomic column at time $\tau$;

            \item $\bar{v}_\gamma(\tau)$ is processed as the candidate derivative of $\gamma$ at the above intersection $\bar{p}(\tau)$, which represents the estimator of the atomic column's drifting direction at time $\tau$ in the TEM application.
        \end{enumerate}
        
        \item Compared to \cite{cheng1999}, our approach uses a simpler building block, the linear functional element \eqref{eq:loc_lin}. To be qualified, \eqref{eq:loc_lin} is constructed to satisfy the local requirements at $t = \tau$, i.e., $\bar{\gamma}_\tau(\tau) = \bar{p}(\tau)$ and $\bar{\gamma}_\tau'(\tau) = \bar{v}_\gamma(\tau).$

        \item The kernel function $K(t)$ and the bandwidth $h$ in \eqref{eq:nonpar} are well studied under relative non-parametric topics like kernel density estimation (KDE) and kernel regression; see \cite{fan2008} for instance.
        
        \item In later practice, we simply use the standard Gaussian kernel and empirically tune the bandwidth, as long as they deliver reasonable performances. Some typical effective choices are $h \in \{.5, 1, 2, \dots\}$.
    \end{enumerate}
\end{rem}

\section{Uncertainty Quantification}\label{sec:uncertainty}
Uncertainty quantification helps understand the accuracy of the recovered $\gamma(t)$, especially at the discrete intersections with every image frame $\Omega(\tau)$. Here we consider only the intermediate estimation $\bar{p}(\tau)$ and proceed with the discretized frame-wise results.

Indeed, if the weights $\Psi(p)$ for $p \in \Omega(\tau)$ are viewed as the probabilities of a discrete distribution supported on $\Omega(\tau)$, then besides the mean $\bar{p}(\tau)$ from \eqref{eq:loc}, the covariance of such distribution could also be empirically calculated as
\begin{equation}\label{eq:cov}
    \bar\Sigma(\tau) = \sum_{p \in \Omega(\tau)} \Psi(p) \big( p - \bar{p}(\tau) \big) \big( p - \bar{p}(\tau) \big)'.
\end{equation}
And hence analogous to the normal distribution, we could utilize the elliptical quadratic form to derive the classical $1 - \alpha$ confidence region of $\bar{p}(\tau)$ as a proxy for that of $\gamma(\tau)$:
$$
CI_\alpha(\tau) = \{p \in \mathbb{R}^3: \big( p - \bar{p}(\tau) \big)' \bar{\Sigma}^+(\tau) \big( p - \bar{p}(\tau) \big) \le Q_{\chi, \alpha} \},
$$
where $Q_{\chi, \alpha}$ is the $(1 - \alpha)$-quantile of the chi-squared distribution with 2 degrees of freedom.

\section{Simulation Study}\label{sec:simu}
To evaluate the algorithm's performance, simulation studies are conducted on image sequences with various ridge (valley) patterns and noise levels. We proceed from a short sequence of images with dimensions $M = N = 41$, $T = 100$, and generate the valley samples similar to the TEM images as
$$
f(m,n;\tau) = C - A(\tau) \cdot \exp \Big( - \frac{\big( m - u(\tau) \big)^2 + \big( n - w(\tau) \big)^2}{2 R(\tau)^2} \Big),
$$
for $(m,n,\tau) \in \Omega$, where constant $C = 140$ is set as the baseline pixel intensities to approximate the vacuum background level of experimental TEM images (e.g., \autoref{fig:tem_eg}), $A(t)$ is the non-negative amplitude function that encodes the valley depths, $\gamma(t) = (u(t), w(t), t)$ is the curve trajectory, and $R(t)$ is the evolving radii function of the (tube-shaped) curve $\gamma(t)$.

Indeed, since the algorithm is designed for potential generalizations, it is necessary to fluctuate the amplitude function $A(t)$. It is finalized to be the continuous combination of trigonometric and constant functions, e.g.,
$$
A(t) = \begin{cases}
    60 & t \in [0, 20) \\
    30 + 30 \cos(\frac{2 \pi t}{10}) & t \in [20, 55) \\
    0 & t \in [55,65) \\
    30 + 30 \cos(\frac{2 \pi t}{10}) & t \in [65, 100)
\end{cases},
$$
where the trigonometric pieces have periodicity $10$. In particular, the period $t \in [55, 65)$ where $A(t) = 0$ aims to simulate the degeneration scenarios when the atomic column is absent in the TEM application. The radius of the curve is also set with some oscillations as $R(\tau) = 6 + 3 \sin(\tau)$.

We also vary the curve trajectory $\gamma(t)$ for simulation completeness. In particular, the following three cases are studied.\begin{enumerate}
    \item Constant, for instance, $\gamma_1(t) = (20, 20, t)$.
    
    \item Discontinuous, for instance, $\gamma_2(t) = (18 + 4 \cdot \mathbbm{1}(t \ge 60), 17 + 6 \cdot \mathbbm{1}(t \ge 30), t)$.
    
    \item Continuously oscillating, for instance, $\gamma_3(t) = (20 + 3 \sin(\frac{2 \pi t}{50}), 20 + 2 \sin(\frac{2 \pi t}{40}), t)$.
\end{enumerate}

Poisson-type noise is applied to the simulated image samples, which is comparable with the integer-valued TEM images' synthetic process \citep{levin2020,manzorro2022}.% And Finally, the image arrays are reversed before inputting into the ridge detection algorithm.

\begin{figure}[htbp] %  figure placement: here, top, bottom, or page
    \centering
    \centerline{\input{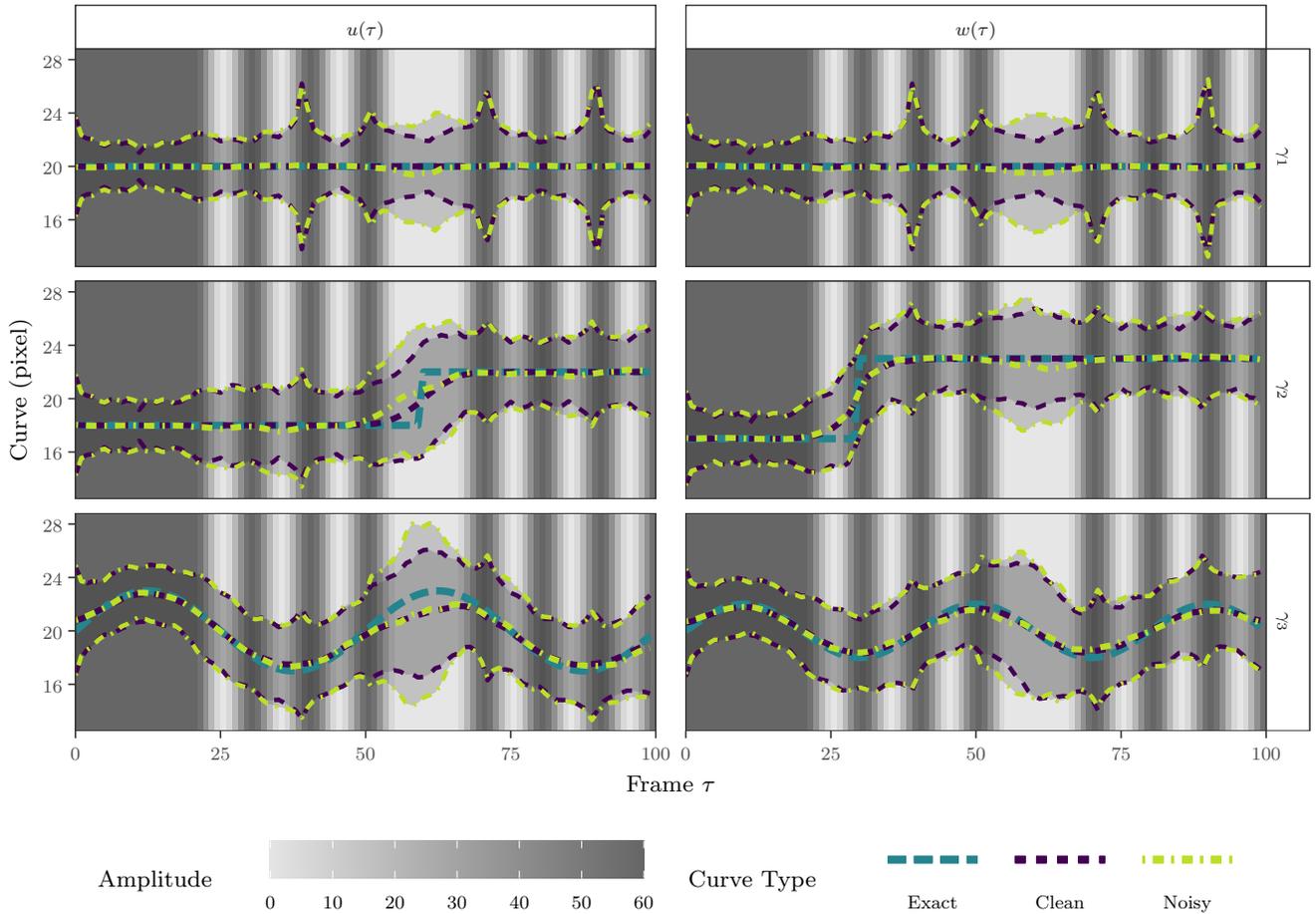}}
    \caption{The detected curves with three underlying ridge definitions are demonstrated with the surrogate marginal confidence regions (shaded bands) given the covariance \eqref{eq:cov}. The darker background indicates more evident valley amplitude. With the analytical `Exact' curves from $\gamma_i$ ($i = 1, 2, 3$) as references, the recovered trajectories from either noise-free images (`Clean') or noisy images (`Noisy') are shown to be close to the truth.}
   \label{fig:simu}
\end{figure}

The simulated noisy video is negated to qualify as the input for ridge detection. The algorithm delivers fair performance for recovering the dynamics of the underlying curvilinear features, though sometimes have limitations under non-continuity or conditions when the pattern completely diminishes for a period of time. Given the simulation results in \autoref{fig:simu}, we are then confident to move forward to the more realistic application of TEM videos. To avoid any implicit potential discrepancy between the physical nanoparticle model and simulated/acquired TEM images, the later analysis will mainly focus on the measurement errors for comparison, i.e., the output differences between the noise-free images and noisy TEM images.

\section{Application}\label{sec:app}

For the TEM applications, the synthetic samples are fed into the algorithm. The resulting performance with respect to the location estimates is compared against other milestone methodologies of the material science community.

\begin{figure}[htbp]
    \captionbox{The baseline nanoparticle structure with the numbers indexing the atomic columns.\label{fig:model}}[.4\textwidth]
        {\includegraphics[width=.4\textwidth]{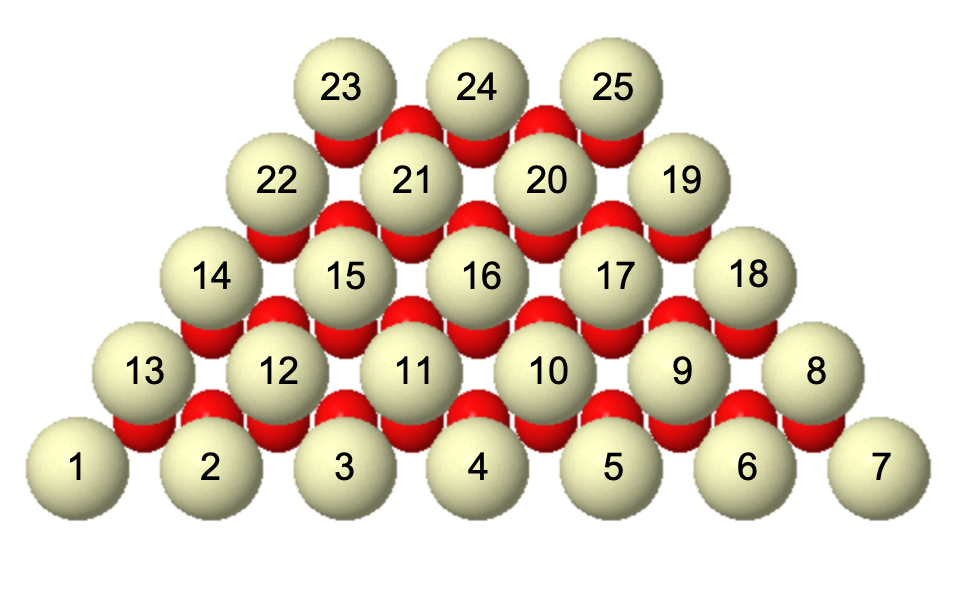}}\hspace{.1\textwidth}%
    \captionbox{The reference noise-free image under the static framework, with the same indexing rule as in \autoref{fig:model}.\label{fig:base_tem}}[.4\textwidth]
        {\includegraphics[width=.4\textwidth]{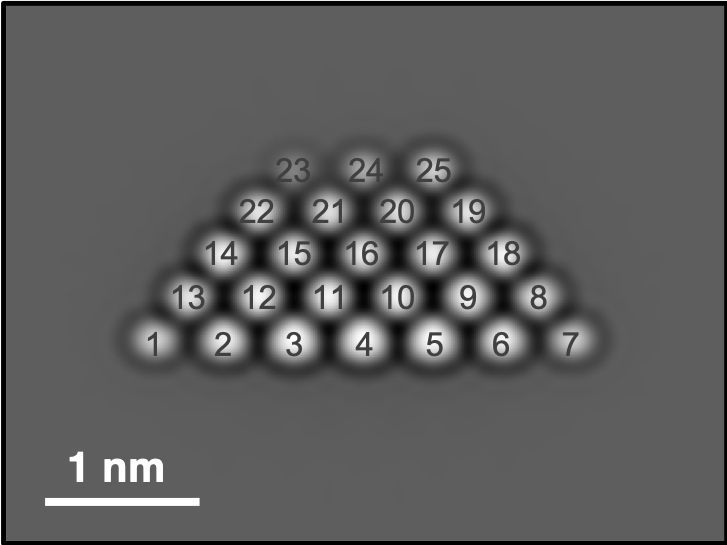}}
\end{figure}

The $\mbox{CeO}_2$ nanoparticle's baseline structure that gets studied is demonstrated as \autoref{fig:model}, while the synthetic samples are extended from it.
% In particular, more details are provided in \autoref{app:syn}.
As mentioned earlier, the major challenge for processing TEM videos comes from the phenomena that some of the atomic columns can have peculiar latent behaviors. Such behaviors frequently lead to the irregular behaviors such as faint contrasts, nonstandard shapes as well as complete absences. Both static and dynamic underlying configurations are studied to evaluate the performance and address the compatibility of our algorithm with those extremes.

\begin{figure}[htbp]
    \centering
    \includegraphics[width=\textwidth]{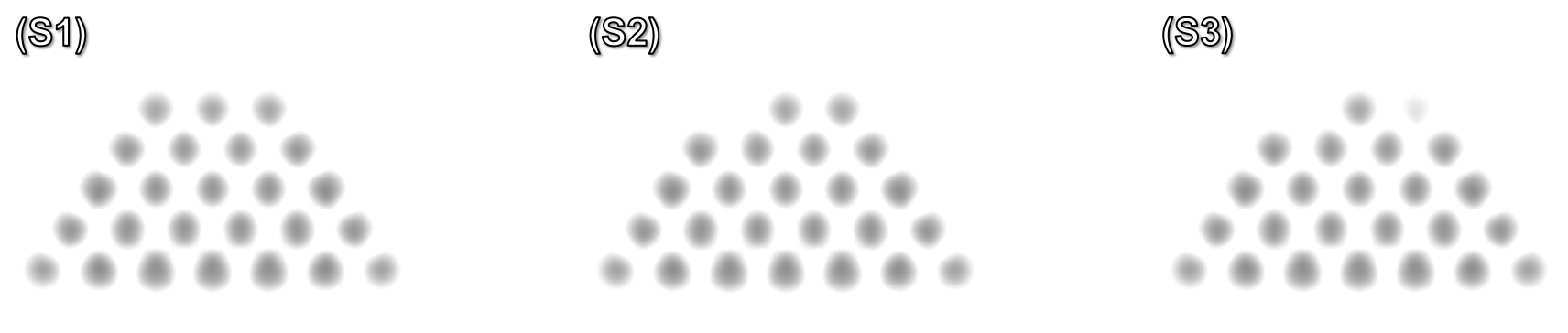}
    \caption{The three candidate configurations for generating 200-frame noisy TEM video, with dynamics.}
    \label{fig:modes}
\end{figure}

\begin{figure}[htbp] %  figure placement: here, top, bottom, or page
    \centering
    \centerline{\input{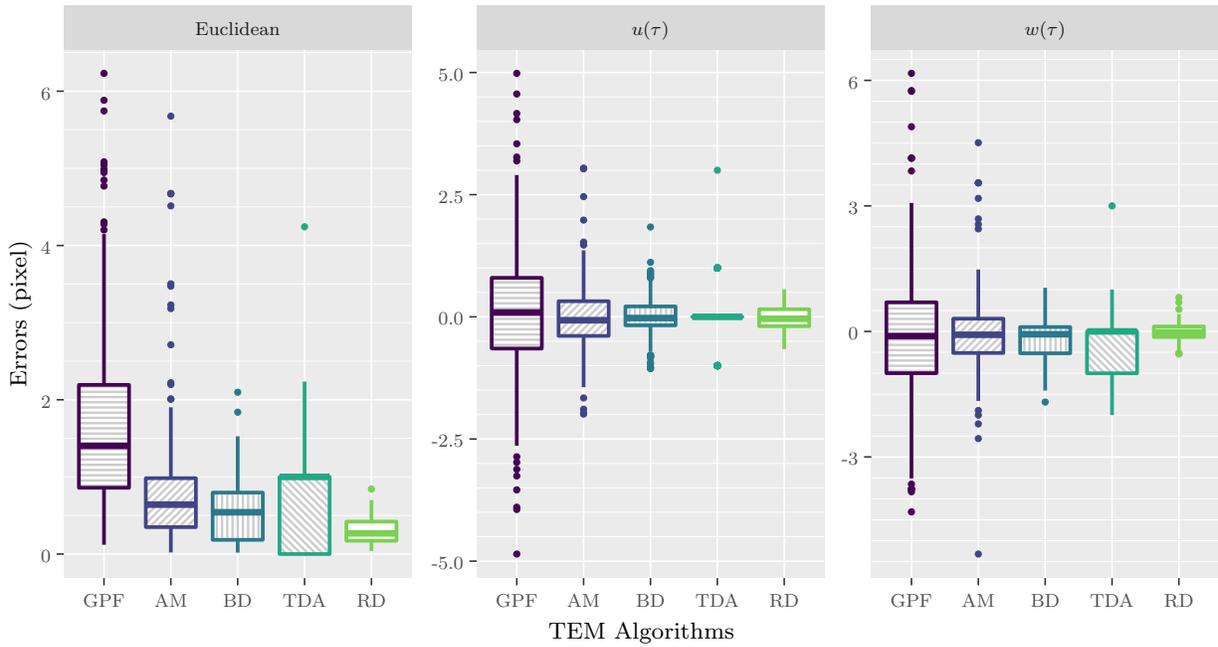}}
    \caption{The error analysis under the static setting. The comparisons are conducted on the two spatial axes as well as the Euclidean errors, between RD and benchmarks including GPF \citep{levin2020}, AM \citep{nord2017}, BD \citep{manzorro2022} and TDA \citep{thomas2022}.}
   \label{fig:static_ebox}
\end{figure}

% \subsection{Synthetic}\label{subsec:syn}
For the static case, based on the noise-free synthetic TEM frame as \autoref{fig:base_tem}, we apply 10 different Poisson noise realizations to generate the image sequence and compare the performance of our proposed algorithm with other benchmarks on both accuracy and consistency.
% The superposition of all the recovered $\check{\gamma}(\tau)$ for $\tau \in [T]$ from our ridge detection algorithm (RD) are demonstrated as in \autoref{fig:static_est}, and such highly concentrated overlap of patterns implies the good estimation stability. For the analysis on a finer scale,
Especially, the error boxes shown in \autoref{fig:static_ebox} indicates that our RD algorithm has the outperforming errors in most aspects if not all.

% \begin{figure}[htbp] %  figure placement: here, top, bottom, or page
%     \centering
%     \centerline{\input{imgs/app_static_est.tikz}}
%     \caption{The superposed recovery output under the static setting. The algorithm is implemented to obtain the locations of all 25 atomic columns from the length-10 synthetic image series with different noise realizations. The characters for each atomic column are completely overlapping given the plot scaling.}
%    \label{fig:static_est}
% \end{figure}

\begin{figure}[htbp] %  figure placement: here, top, bottom, or page
    \centering
    \centerline{\input{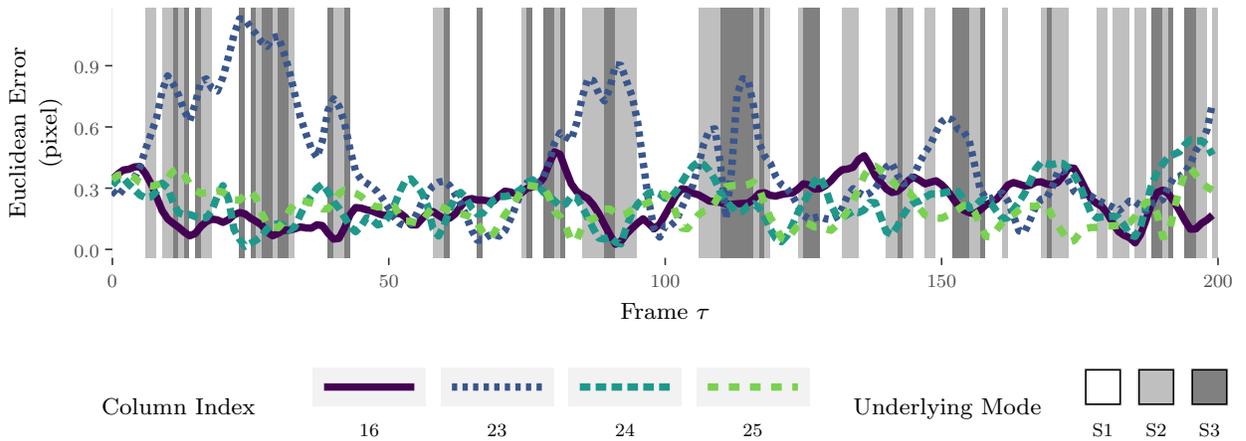}}
    \caption{The algorithm performance under the dynamic setting. The three different underlying configurations (\autoref{fig:modes}) are coded by the background gray-scale levels. The fitted Euclidean deviation of selective atomic columns, from both interior (column 16) and surface (columns 23, 24, 25), are plotted for complete comparisons.}
   \label{fig:dynamic}
\end{figure}

We then proceed to the dynamic case and construct the TEM video of nanoparticles with the three structural configurations in \autoref{fig:modes}. The series of the underlying configurations is obtained by simulating a Markov process, with a manually specified transition probability matrix
$$
\begin{blockarray}{cccc}
     & \mbox{To S1} & \mbox{To S2} & \mbox{To S3} \\
    \begin{block}{c(ccc)}
    \mbox{From S1} & 0.85 & 0.10 & 0.05 \\
    \mbox{From S2} & 0.30 & 0.55 & 0.15 \\
    \mbox{From S3} & 0.25 & 0.30 & 0.45 \\
    \end{block}
\end{blockarray}.
$$
The surface atomic columns indexed as 23 and 25 may experience either absences or intensity changes during the video. We fix a simulated realization of the underlying configuration series and implement the algorithm on both noise-free and noisy images. The Euclidean distances between the recovered ridge points $\check\gamma(\tau)$ ($\tau \in [T]$) from the two sets of outputs are analyzed. With the discrete pixel resolutions of images, the results shown in \autoref{fig:dynamic} are satisfactory, especially given that the errors for the most volatile column 23 are almost always below unit-pixel.

% \subsection{Experimental}\label{subsec:exp}
% Since there is no ground truth behind such raw experimental samples, the performance of the algorithm is presented by comparison with other benchmarks. 

\section{Conclusion}\label{sec:con}

In this work, we proposed the non-parametric approach to continuously recover the ridge pattern from (TEM) videos, provided the ridge is parameterized temporally. Our algorithm explored the geometric local properties as well as the continuity restraint, and established the scores for pixels being on the ridge. The kernel-based functional estimator is used to output the ridge curve. We tailored the algorithm specially for the TEM application to tackle the disappearing and re-appearing atomic columns. We also included analysis for uncertainty quantification. Finally we evaluated our algorithm with carefully designed simulations, and implemented it on the synthetic image examples for TEM applications.

In addition to the synthetic material science applications, the algorithm may accomplish its most promising value when implemented on some real raw TEM images. Currently, the approach still has limited generalizability as it only recovers the ridge (or valley) curves with some edge degeneration scenarios. Other future directions include extending the applicability to detect connector curves, i.e., the intermediate segments between ridges and valleys \citep{damon1999}. Under the time series analysis or the signal processing framework, the extension to connector curves can be useful to extract information propagation patterns, especially in topics such as impulse response analysis and change point detection.

%%%%%%%%%%%%%%%%%%%%%%%%%%%%%%%%%%%%%%%%%%%%%%
%% Single Appendix:                         %%
%%%%%%%%%%%%%%%%%%%%%%%%%%%%%%%%%%%%%%%%%%%%%%
\begin{appendix}
\section{Gradient-yielded Alternative Approximations}\label{app:grad}
Here we introduce an alternative analysis using gradient-yielded approximations as an optional supplementary update to the intermediate steps of the approach developed in the main paper. Some additional definitions and relative corollaries are summarized. Their successive remarks elaborate more details about the strengths of such gradient-yielded approximations qualitatively.

To give an overview, the introduction of the gradient-yielded quantities from \autoref{cor:approx} serves as the primary delivery of this section. These quantities have especially strong estimation power at the pixels that are within a neighborhood of non-degenerated ridge segments. Specifically, we will call those pixels as \emph{non-degenerated neighbor pixels} in the following context for conciseness.

We start with the following definition introducing the normalized gradient vectors.

\begin{defn}\label{def:newton}
    Define the normalization of the gradient vectors $\widetilde{\nabla}(p)$ as
    \begin{equation}\label{eq:normGrad}
        \widehat{\nabla}(p) = \widetilde{\Delta}^+(p) \widetilde{\nabla}(p).
    \end{equation}
\end{defn}

\begin{rem}\label{rem:newton}
\begin{enumerate}[(a)]
    \item The normalization is motivated by the Newton's method from optimization \citep{nocedal2006}. It can help robustify the estimated gradient directions at non-degenerated neighbor pixels; see \Cref{rem:features,rem:updateweight} for more detailed arguments.

    \item For any $p$ among the non-degenerated neighbor pixels, the (estimated) Hessian matrix is usually invertible. Furthermore, the \autoref{enum:Tang} of \Cref{def:ridge} also guarantees that $\widehat{\nabla}(p)$ as the approximation of $\Delta^+(p) \nabla(p)$ is nearly parallel to $v_{1}(p)$.
\end{enumerate}
\end{rem}

Given the normalized gradient estimators by \autoref{def:newton}, we can obtain the following corollary that introduces approximations to the quantities proposed in \autoref{prop:approx} at those non-degenerated neighbor pixels. The proof can be found in \autoref{app:proof}.

\begin{cor}\label{cor:approx}
If a non-degenerated segment of the ridge $\gamma$ passes near a lattice point $p \in \Omega$, then
\begin{enumerate}
    \item With $\widehat{\nabla}(p)$ redirected such that $\big< \widehat{\nabla}(p), e_t \big> \ge 0$,
    \begin{equation}\label{eq:vhat}
        \hat{v}(p) = \frac{\widehat{\nabla}(p)}{\|\widehat{\nabla}(p)\|} \approx v(p).
    \end{equation}
    
    \item 
    \begin{equation}\label{eq:lambdahat}
        \hat{\lambda}(p) = \argmin_{\ell \in \mathbb{R}} \| \widetilde{\Delta}(p) \hat{v}(p) - \ell \hat{v}(p)\| = \hat{v}'(p) \widetilde{\Delta}(p) \hat{v}(p) \approx \lambda(p).
    \end{equation}

    \item 
    \begin{equation}\label{eq:rhohat}
        \hat{\rho}(p) = \big| \cos\big( \hat{v}(p), e_t \big) \big|\approx \rho(p).
    \end{equation}

    \item 
    \begin{equation}\label{eq:thetahat}
        \hat\theta(p) = \Big| \cos\big( \hat{v}(p), \widetilde{\nabla}(p) \big) \Big| \approx \theta(p).
    \end{equation}
    
    \item 
    \begin{equation}\label{eq:kappahat}
    \begin{dcases}
        \hat{\eta}(p) = \frac{2 \hat{\xi}_2(p)}{\hat{\xi}_1^2(p) - 2 \hat{\xi}_2(p)} \approx \eta(p)\\
        \hat{\kappa}(p) = \hat{\xi}_2(p) - \hat{\lambda}^2(p) \approx \kappa(p)
    \end{dcases}
    \end{equation}
    where
    $$
    \begin{aligned}
        \hat{\xi}_1(p) & = \trace\big[ \widetilde{\Delta}(p) \big] - \hat{\lambda}(p) \approx \lambda_{2}(p) + \lambda_{3}(p),\\
        \hat{\xi}_2(p) & = \frac{1}{2} \Big[ \hat{\xi}^2_1(p) - \big( \|\widetilde{\Delta}(p)\|_F^2 - \hat{\lambda}^2(p) \big) \Big] \approx \lambda_{2}(p) \lambda_{3}(p).
    \end{aligned}
    $$
\end{enumerate}
\end{cor}

\begin{rem}\label{rem:features}
    The hatted notations in \Cref{cor:approx}, e.g., $\hat{v}(p)$ and $\hat{\lambda}(p)$, always represent the gradient-yielded approximations of the eigen-related terms. For conciseness, we may not include similar corollaries for every propositions or definitions in the main paper, but the gradient-yielded counterparts for those statements can be straightforwardly derived and the notations of relative terms analogously inherit the upper hats. The two versions of quantities possess distinct features and emphasize different perspectives in practice:\begin{enumerate}[(a)]
        \item The unhatted quantities, or those originally given in \autoref{prop:approx}, manipulate the empirical eigen-decomposition of the scaled Hessian estimators \eqref{eq:updateHess} and aim for universal applicability. They are especially effective under degenerated scenarios. Nevertheless, they are usually accompanied with moderate performance around the non-degenerated ridge segments.

        \item The gradient-yielded approximations for the eigen-terms, as proposed in \autoref{cor:approx}, augment the concentration of the curve's local directions. Specifically, these approximations will introduce a stronger force that navigates the pixels around non-degenerated ridge segments towards the true curve within a certain neighborhood. Analogous to \autoref{rem:newton} and the arguments for developing the Newton's method in optimization, it is due to the underlying landscape structure of the mapping.
    \end{enumerate}
\end{rem}

We can then define the gradient-yielded intra-frame metrics analogous to \autoref{def:intra} with hatted approximations from \autoref{cor:approx}, and propose the following corollary as an enhanced update of \autoref{def:weight}.

\begin{cor}\label{cor:weight}
Given a frame $\Omega(\tau)$ for $\tau \in [T]$, the gradient-yielded counterparts of the metrics in \Cref{def:intra} give the approximated weight for $p \in \Omega(\tau)$
\begin{equation*}\label{eq:weighthat}
    \hat\Phi(p) = \exp\big( \hat L_\rho(p) + \hat L_\theta(p) + \hat L_{\eta,\kappa}(p) \big).
\end{equation*}
And consequently update the weight and local direction by
\begin{align}
    v(p) & \mapsto \frac{\Phi(p) v(p) + \hat\Phi(p) \hat{v}(p)}{\Phi(p) + \hat{\Phi}(p)}.\label{eq:vbar}\\
    \Phi(p) & \mapsto \Phi(p) + \hat{\Phi}(p),\label{eq:weightFinal}
\end{align}
\end{cor}

\begin{rem}\label{rem:updateweight}
    \begin{enumerate}[(a)]
        \item The \autoref{cor:weight} merges the (hatted) gradient-yielded metric designs with the (unhatted) \autoref{def:weight}, and combines the featured strength of the two systems listed in \autoref{rem:features} for the enhanced estimation. In particular, the (unhatted) terms $v(p)$ and $\Phi(p)$ will supposedly dominate in \eqref{eq:vbar} and \eqref{eq:weightFinal} under degeneration scenarios, while the hatted ones will play more essential roles when considering the non-degenerated neighbor pixels.

        \item The update \eqref{eq:vbar} partially inherits the strengths of the gradient-yielded approximation $\hat{v}(p)$ as mentioned in \autoref{rem:features}, especially at the non-degenerated neighbor pixels. In particular, the updated vector is likely to point towards the ridge curve due to the mapping's gradient landscape.

        \item If applicable, the update \eqref{eq:vbar} of $v(p)$ impacts the candidate ridge tangent $v_\gamma(p)$ \eqref{eq:candidate}, which is used in the roughness penalization (\autoref{def:rough}) and the curve connection (\autoref{def:nonpar}). Similarly, the update \eqref{eq:weightFinal} of $\Phi(p)$ affects the inter-frame weight (\autoref{def:score}), the curve connection (\autoref{def:nonpar}) and the uncertainty statements in \autoref{sec:uncertainty}.
    \end{enumerate}
\end{rem}

% \section{Details on the synthetic TEM model}\label{app:syn}
% The nano-particle structure provided in \autoref{fig:model} is summarized from an experimental TEM image sequence illustrated in \autoref{fig:tem_eg}.
% 
% Extending from the physics model in \autoref{fig:model}, the synthetic static TEM image \autoref{fig:base_tem} is generated.
% 
% In addition, by varying the atom numbers of some atomic columns in \autoref{fig:model}, \autoref{fig:modes} can be created.
% {\color{red} PETER, please help include some physics introduction about the experimental TEM sequence in \autoref{fig:tem_eg}, the nanoparticle model in \autoref{fig:model} and the mechanism to generate the synthetic TEM images \Cref{fig:base_tem,fig:modes}. Please feel free to edit the structure of this section and revise any written sentences.}

\section{Proof}\label{app:proof}
Herein, we show detailed justification for some selected statements that appear throughout this paper.

\begin{proof}[Proof of \autoref{prop:approx} \& \autoref{cor:approx}]
For this proof we always consider the limiting condition that the noise level is approaching zero and the convolution effect of derivative calculation tends to vanish. For simplicity, we only show the ordinary convergence results. The extension to stochastic convergence is straightforward.

First of all, if a grid point $p$ is exactly on the ridge, the derivation of the proof is trivial from \autoref{def:ridge}.

Instead, suppose the grid point $p \in \Omega$ is relatively close to a ridge point $q \in \gamma$,
% there exists a ridge point $q \in \gamma$ such that its small neighborhood contains the point $p \in \Omega$
and we will prove that the approximations are of high quality. In addition, here the eigen-related quantities for point $q$, i.e. $\{\lambda_1(q), \lambda_2(q), \lambda_3(q), v_1(q), v_2(q), v_3(q)\}$, always refer to those analytically decomposed from the exact Hessian $\Delta(q)$. Since eigen-decomposition on the Hessian matrix of a differentiable mapping is also continuous, the estimations' approximation quality of the unhatted terms can be directly guaranteed using continuity arguments. Hence we will mainly focus on the behaviors of the gradient-yielded (hatted) quantities in \autoref{cor:approx} where $\Delta(q)$ and $\nabla(q)$ are both non-singular.

Without loss of generality, we can assume that the unit length vector $\hat{v}(p) = \frac{\widehat{\nabla}(p)}{\| \widehat{\nabla}(p) \|}$ can be expressed as the linear combination
$$
\hat{v}(p) = \sum_{i = 1}^3 k_i v_{i}(p), \qquad \mbox{where} \qquad \sum_{i=1}^3 k_i^2 = 1.
$$
Note that $\widehat{\nabla}(p) = \widetilde{\Delta}^+(p) \widetilde{\nabla}(p)$ well approximates $\Delta^+(q) \nabla(q) ~ \propto ~ v_{1}(q)$ given that $q \in \gamma$ and $p$ is close to $q$.
\begin{enumerate}
    \item Since $\hat{v}(p) = \frac{\widehat{\nabla}(p)}{\|\widehat{\nabla}(p)\|}$ is continuous with respect to non-singular $\widehat{\nabla}(p)$, and given that $\widehat{\nabla}(p) \approx \| \Delta^+(q) \nabla(q) \| \cdot v_1(q)$, then continuity implies that $\hat{v}(p) \approx v_1(q)$.
    % the approximation statement $\hat{v}(p) \approx v(p)$ follows by continuity. And subsequently with $v(p) \approx v_{1}(q)$, we have $\hat{v}(p) \approx v_{1}(q)$.

    \item The statements on $\hat{\rho}(p)$ can be concluded directly by definition of cosine similarities.

    \item From the conclusions above, we have that $k_1 = \big< \hat{v}(p), v_1(p) \big> \approx \big< v_1(q), v_1(p) \big> \approx 1$, $k_2 \approx 0$ and $k_3 \approx 0$. Since $\hat{\lambda}(p) = \sum_{i=1}^3 k_i^2 \lambda_{i}(p)$ is continuous with respect to $k_i$, and $\lambda_{1}(p) \approx \lambda_{1}(q)$, we conclude that $\hat{\lambda}(p) \approx \lambda_{1}(q)$.

    \item It directly follows from the first point above given continuity.

    \item It suffices to show that $\hat{\xi}_1(p) \approx \lambda_{2}(q) + \lambda_{3}(q)$ and $\hat{\xi}_2(p) \approx \lambda_{2}(q) \lambda_{3}(q)$. Some calculations give
    $$
    \hat{\xi}_1(p) = \sum_{i=1}^3 (1 - k_i) \lambda_{i}(p), \qquad \hat{\xi}_2(p) = \lambda_{2}(p) \lambda_{3}(p) + \big( \lambda_{1}(p) - \hat{\lambda}(p) \big) \big( \lambda_{2}(p) + \lambda_{3}(p) - \hat{\lambda}(p) \big).
    $$
    Then the convergence statements follow by continuity again.
\end{enumerate}

Under degeneration, on the other hand, the gradient-yielded approximations can perform rather poorly, but they still give bounded quantities and will be compensated by the original unhatted terms in \autoref{cor:weight} based on our algorithm designs.
\end{proof}

\begin{proof}[Proof of \autoref{prop:intra_behavior}]
\begin{enumerate}
    \item The monotonicity statement is true according to the definition of cosine similarity. Since $\max L_\rho(p) = 2$, then $L_\rho(p) = 2 \Leftrightarrow \rho(p) = 1 \Leftrightarrow v(p)$ is parallel to the temporal indicator $e_t$.

    \item The monotonicity statement is true according to the definition of cosine similarity. Since $\max L_\theta(p) = 2$, then $L_\theta(p) = 2 \Leftrightarrow \theta(p) = 1 \Leftrightarrow v(p)$ is parallel to $\widetilde{\Delta}(p) \widehat{\nabla}(p) = \widetilde{\nabla}(p)$.

    \item It can be shown with simple algebra that
    $$
    \frac{2}{\eta(p)} = \frac{\lambda_{2}(p)}{\lambda_{3}(p)} + \frac{\lambda_{3}(p)}{\lambda_{2}(p)} \in (-\infty, -2] ~ \bigcup ~ [2, \infty).
    $$
    It implies that $\eta(p)$ is positive only when $\lambda_{2}(p)$ and $\lambda_{3}(p)$ have the same sign, and the maximum $\eta(p) = 1$ is attained if and only if $\lambda_{2}(p) / \lambda_{3}(p) = 1$. In addition, since $L_{\eta, \kappa}(p)$ is monotonic with respect to both $\eta(p)$ and the product $\lambda_{2}(p) \lambda_{3}(p)$ due to monotonicity-conserved composition with $\kappa(p)$, respectively, the sub-statement holds.
\end{enumerate}

Similar derivations for the gradient-yielded version can be analogously obtained.
\end{proof}
\end{appendix}

%%%%%%%%%%%%%%%%%%%%%%%%%%%%%%%%%%%%%%%%%%%%%%
%% Support information, if any,             %%
%% should be provided in the                %%
%% Acknowledgements section.                %%
%%%%%%%%%%%%%%%%%%%%%%%%%%%%%%%%%%%%%%%%%%%%%%
\begin{acks}[Acknowledgments]
The authors thank Ramon Manzorro (University of Cádiz) and Professor Peter A. Crozier's lab members for assistance in imaging simulations, data handling and material science background education.
% The authors thank the anonymous Reviewer and the Associate editor for their comments which helped improving the paper. 
\end{acks}
%%%%%%%%%%%%%%%%%%%%%%%%%%%%%%%%%%%%%%%%%%%%%%
%% Funding information, if any,             %%
%% should be provided in the                %%
%% funding section.                         %%
%%%%%%%%%%%%%%%%%%%%%%%%%%%%%%%%%%%%%%%%%%%%%%
\begin{funding}
The authors gratefully acknowledge financial support from National Science Foundation Awards 2114143, 1934985, 1940124, and 1940276.
% This research was conducted with support from the Cornell University Center for Advanced Computing, which receives funding from Cornell University, the National Science Foundation, and members of its Partner Program.
\end{funding}
\bibliographystyle{imsart-nameyear} % Style BST file (imsart-number.bst or imsart-nameyear.bst)
\bibliography{bibs/Algo,bibs/App,bibs/Bkgrnd,bibs/LCS, bibs/Nonpar,bibs/TEM_algo,bibs/TEM_bg}
% Bibliography file (usually '*.bib')

%% or include bibliography directly:
% \begin{thebibliography}{}
% \bibitem{b1}
% \end{thebibliography}

\end{document}